\documentclass[
 reprint,
superscriptaddress,
 amsmath,amssymb,
 aps,
]{revtex4-2}
\usepackage[utf8]{inputenc}
\usepackage[T1]{fontenc}
\usepackage{lineno}
\usepackage{graphicx}
\preprint{APS/123-QED}
\usepackage[dvipsnames]{xcolor}

\begin{document}
\title{Conveyor-mode single-electron shuttling in Si/SiGe for a scalable quantum computing architecture}

\author{Inga Seidler}
\author{Tom Struck}
\author{Ran Xue}
\author{Niels Focke}\affiliation{JARA-FIT Institute for Quantum Information, Forschungszentrum J\"ulich GmbH and RWTH Aachen University, Aachen, Germany}
\author{Stefan Trellenkamp}
\affiliation{Helmholtz Nano Facility (HNF), Forschungszentrum J\"ulich, J\"ulich, Germany}
\author{Hendrik Bluhm}
\author{Lars R. Schreiber}\affiliation{JARA-FIT Institute for Quantum Information, Forschungszentrum J\"ulich GmbH and RWTH Aachen University, Aachen, Germany}

\begin{abstract}
Small spin-qubit registers defined by single electrons confined in Si/SiGe quantum dots operate successfully and connecting these would permit scalable quantum computation. Shuttling the qubit carrying electrons between registers is a natural choice for high-fidelity coherent links provided the overhead of control signals stays moderate. Our proof-of-principle demonstrates shuttling of a single electron by a propagating wave-potential in an electrostatically defined 420\,nm long Si/SiGe quantum-channel. This conveyor-mode shuttling approach requires independent from its length only four sinusoidal control signals. We discuss the tuning of the signal parameters, detect the smoothness of the electron motion enabling the mapping of potential disorder and observe a high single-electron shuttling fidelity of $99.42\pm0.02\,\%$ including a reversal of direction. Our shuttling device can be readily embedded in industrial fabrication of Si/SiGe qubit chips and paves the way to solving the signal-fanout problem for a fully scalable semiconductor quantum-computing architecture.
\end{abstract}

\flushbottom
\maketitle

As single electron-spin qubits confined in electrostatically defined Si/SiGe quantum dots (QDs) have overcome the fidelity threshold for quantum error correction for both single and two-qubit gates \cite{Yoneda2018,Zajac2018,Watson2018,Xue2019,Xue2021} and high-fidelity qubit readout has become accessible \cite{Connors2020,Noiri2020, Kammerloher2021}, the research focus has been moving towards scalable quantum computing architectures \cite{Hollenberg2006,Vandersypen2017,Veldhorst2017,Li2018,Boter2019}. A crucial element for scalability is a coherent qubit coupling-mechanism on a medium (1 to 10 micron) and long-range distance (> 1\,mm), to connect dense qubit registers and overcome the signal-fanout problem, i.e. fan out the signal lines from densely packed electrostatic gates to control electronics  \cite{Vandersypen2017}. Coupling via spin-to-photon-to-spin conversion by transferring the spin information to a cavity mode proofs promising as a long range coupler \cite{Mi2018,Samkharadze2018,Landig2018,Borjans2020,Borjans2020_2}. Besides a micromagnet, it requires a high quality superconducting cavity. The fabrication of which is critically compatible to fabrication processes necessary for the gate electrodes defining the QDs. Shuttling the electron qubit itself towards another (static) qubit is promising approach for spin-coherent medium range coupling, since it employs the local exchange-based high-fidelity two-qubit gate and suppresses any residual coupling to other distant qubits. Electron transfer and entanglement of two separately transferred spins shuttled by a surface acoustic wave has been shown for GaAs/(Al,Ga)As based devices \cite{Hermelin2011,McNeil2011,Takada2019,Jadot2021}. The transfer velocity is set by the material's sound velocity, thus ac-/decelerating the qubit without uncontrolled orbital excitation is involved \cite{Bertrand2016} and the implementation to Si/SiGe requires an additional proximal material with high piezoelectricity \cite{Du2007}. Controlling the electron shuttling by an array of metal gates is thus a natural choice and has been demonstrated \cite{Baart2016,Flentje2017,Mills2019,Yoneda2021}. In all these demonstrations, electrons are shuttled via a series of Landau-Zener transitions through a one-dimensional array of tunnel-coupled QDs (termed bucket-brigade mode shuttling) \cite{Zhao2019,Buonacorsi2020,Kryzwda2020,Ginzel2020,Krzywda2021}, the tunnel-coupling and chemical potential of which need to be carefully tuned by the applied voltages. The longer the shuttling device, the more input signals are required increasing the tuning complexity. Thus, this is a debatable solution to the signal-fanout problem limiting scalability.

In this manuscript, we present a proof-of-principle demonstration of a new mode of single electron shuttling termed conveyor mode in a shuttling device named quantum bus (QuBus). Conveyor-mode shuttling is based on four input signals which form a propagating sinusoidal potential. The electron is transported smoothly and adiabatically in one of the pockets of the propagating wave confined to an undoped Si/SiGe heterostructure. In contrast to other approaches, the number of gates required for a QuBus device is thus independent from its length and effort for tuning is largely reduced, thus potentially solving the signal fanout problem. The velocity, acceleration and transfer distance can be adjusted by changing the frequency and duration of the input signals. Furthermore, the fabrication of the QuBus is technologically compatible with the fabrication of the electrostatically defined QD-devices. We show smooth shuttling by time-resolved tracking of the electron motion with shuttling fidelities above 99\%.

\begin{figure}
    \centering
    \includegraphics[width=\linewidth]{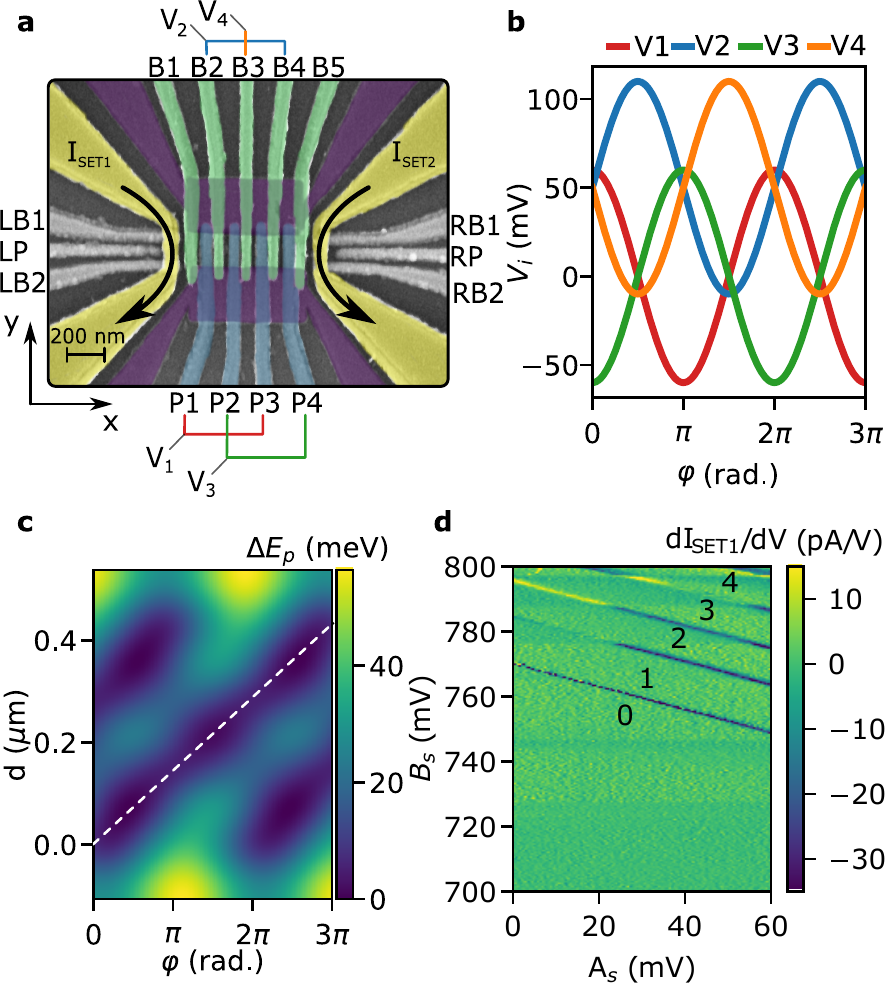}
    \caption{Device layout and shuttling pulse. \textbf{a}  Coloured scanning electron micrograph (SEM) of the three-gate layer design used for the device. Lines indicate electrical connections among clavier gates labelled $B_i$ and $P_i$. $V_i(t)$ labels the voltage signal trace applied to electrically connected gate sets. The accumulation gate of each SET is marked in yellow. \textbf{b} Voltage traces $V_i(t)$ applied to the electrically connected gate sets as indicated in panel a during electron shuttling plotted as a function of $\varphi(t)$ where we here set $A_s=60\,$mV, $B_s=0\,$V and $\Delta B_s=50$\,mV. \textbf{c} Electrostatic simulation of the potential difference $\Delta E_p$ in the strained Si quantum well underneath the clavier gates, if only the voltages $V_i(t)$ applied to the clavier gates according to panel b are taken into account. The position of the shuttled QD is indicated by a dotted white line. The simulation uses an offset $\Delta B_s$ between the gate layers of 50\,mV, a shuttling amplitude $A_s$ of 60\,mV and a Voltage of -150\,mV is applied to the screening gates. \textbf{d} Charge diagram where the amplitude of the shuttling pulse is varied on the x-axis and the total voltage offset on the channel gates is changed along the y-axis. The numbers in the image indicate the electron filling at that point.}
    \label{fig:fig1}
\end{figure}

Before the proof-of-principle demonstration of electron shuttling, we explain the concept of conveyor-mode shuttling in our QuBus device. The device  consists of three patterned metal-gate layers fabricated on top of a planar Si/SiGe heterostructure (see Methods for details). The long screening gates in the lowest metal layer (colored purple in Fig. \ref{fig:fig1}a) are kept at 0\,V throughout the entire measurement and form a one-dimension electron channel (1DEC) 
along the x-direction in the Si/SiGe quantum well. At each end of the 1DEC, a single electron transistor (SET) is induced by accumulation gates (colored yellow purple in Fig. \ref{fig:fig1}a), barrier gates (LB1, LB2, RB1, RB2) and plunger gates (LP and RP). The SET serves a dual purpose as proximal charge detector and electron reservoir tunnel-coupled to the 1DEC. The B and P gates on top of the 1DEC could either form up to four QDs (barriers B$_i$, plungers P$_i$), but here they are used to create travelling wave potential in the 1DEC and are referred to as the clavier gates. Each fourth gate is electrically connected as indicated by the labels $V_1, V_2, V_3,$ and $ V_4$ in Fig. \ref{fig:fig1}a. The gates $B_1$ and $B_5$ are each connected separately to control the tunnel barrier between the each end of 1DEC and the SET. For shuttling, we apply a simple sine voltage to the clavier gates (Fig. \ref{fig:fig1}b):
\begin{equation}
V_i = A_s \cos(\varphi(t)-\pi/2(i-1))+B_s+\Delta B_s (i\mod2)
\end{equation}
where $i = 1\dots4$ and the phase is given by $\varphi(t) = 2\pi f\cdot t $ with $f$ and $t$ being the shuttling frequency and shuttling time, respectively. $A_s$ is the common amplitude of the sine waves, $B_s$ and $\Delta B_s$ are the dc voltage offset on all the clavier gates and the additional offset used for $V_2$ and $V_4$, respectively. 

This signal creates a travelling wave potential in the 1DEC with a smoothly propagating QD moving from the left side, at $d=0\,$nm, to the right side, at $d=420\,$nm, corresponding to a $\phi_{max}=3\pi$, marked by the white dashed line in Fig. \ref{fig:fig1}c. The pitch of the clavier gates $g=70$\,nm determines the wavelength $\lambda=4g=280\,$nm of the travelling wave potential in the 1DEC. Note that we need to a apply a different voltage offset $\Delta B_s$ to compensate for their different lever arms to the 1DEC, since the P- and B-gates are fabricated on the second and third metal layer, respectively. 

To initiate electron shuttling, the 1DEC is depleted by a flush pulse sequence (see Methods) and one single electron is then loaded from an SET, tuned by $B_S$ and $A_S$ at $\varphi=0$ (charge diagram in Fig. \ref{fig:fig1}b for left SET), where $B_S$ sets the overall chemical potential within the 1DEC and $A_S$ the confinement of the QD formed at the left end of the 1DEC. For measurement of charge occupation, the tunnel barrier between SET and 1DEC is transparent (set by the voltage on gate B$_1$). After loading an electron, the tunnel barrier is set opaque completing the initialization of the QuBus. The charge diagram in Fig. \ref{fig:fig1}d and tuning of the tunnel barriers to the reservoirs with similar charge diagrams poses the only voltage tuning required for our QuBus.

For the proof-of-principle demonstration, we implement two shuttling pulse sequences: (I) Shuttling forth and back starting from the left hand side of the 1DEC. (II) Shuttling through the 1DEC from left to right hand side. We label the different pulse segments in the following way: S$_n$ terms a shuttling pulse segment moving the electron by a distance of $ n \lambda$ along the x-direction. P$_n$ and D$_n$ are charge preparation and detection pulse segments with index $n$ being L and R for left or right side of the QuBus, respectively. Measuring the sensor currents throughout each pulse sequence allows us to obtain single-shot time-resolved data.

\begin{figure*}
    \centering
    \includegraphics[width=\linewidth]{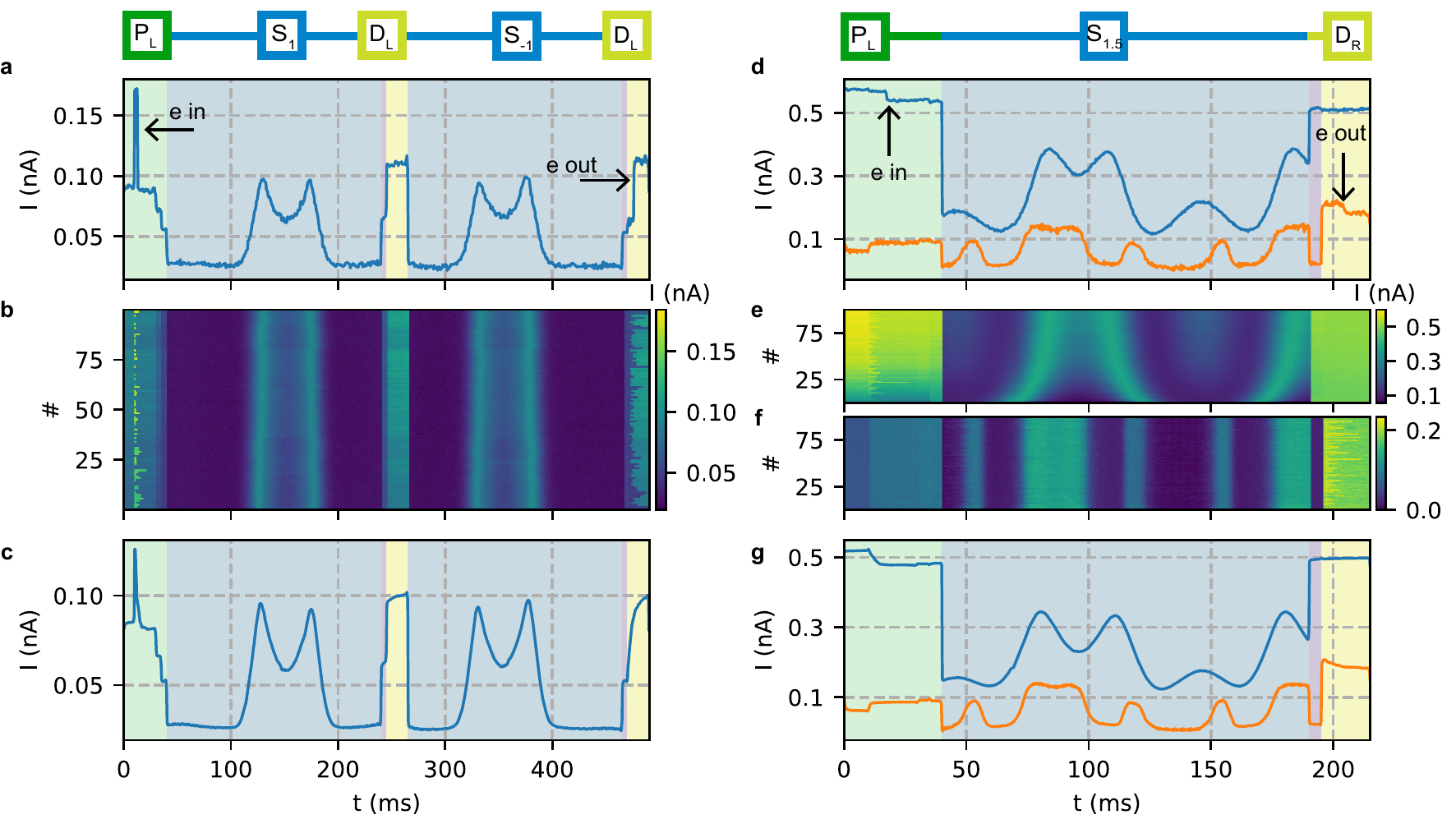}
    \caption{Sensor signal traces for two shuttling pulse sequences. Firstly, a shuttling back and forth in the device is realized by loading an electron on the left, shuttling a distance of  $\lambda$ to the right, detecting on the left, shuttling a distance  of $\lambda$ to the left and detecting on the left (a-c). The second pulse sequence shuttling through the device consists of loading on the left, shuttling 1.5$\lambda$ to the right and detecting on the right (d-g). The pulse sequences are indicated above the panels. \textbf{a} Single trace of the sensor response for shuttling in and out on the left. The plot background color marks the duration of each pulse segment. \textbf{b} Left sensor signal of 100 shuttling traces. \textbf{c} Average sensor response of the 100 shuttling traces in panel b. \textbf{d} Single trace of left (blue) and right (orange) sensor response for shuttling through the device. \textbf{e} Left sensor signal of 100 shuttling traces. The current background change is due to automatic tuning of the sensor set point. \textbf{f} Right sensor signal of the same 100 shuttling traces. \textbf{g} Average sensor response of the 100 shuttling traces depicted in panels e,f. Both pulse sequences are applied with $A_S=50\,$mV, $B_S=730\,$mV and $\Delta B_S=100\,$mV.}
    \label{fig:fig2}
\end{figure*}

For the first pulses sequence, we assemble five different pulse segments: P$_L$,S$_1$,D$_L$,S$_{-1}$,D$_L$ (Fig. \ref{fig:fig2}a). During P$_L$, (marked green in Fig. \ref{fig:fig2}a), a single electron is loaded from the left SET, as detected by the signal step of the left SET current from 0.17$\,$nA down to 0.09$\,$nA, since charging the 1DECs slightly modifies the operation point of the SET's I-V characteristic. After loading the electron, the tunnel barrier to the SET is raised again to complete the initialization. During the first shuttling segment S$_1$ indicated in blue, the electron is moved to the right (positive x-direction) by the distance $1\lambda$. To check whether the electron has moved, a detection segment D$_L$, colored in yellow, is included. The second shuttling segment S$_{-1}$ transfers the electron back from the right to the left (negative x-direction) by $1\lambda$. The second detection segment D$_L$ determines whether the electron has returned to its initial position. The short segment marked in purple in Fig. \ref{fig:fig2}a prepares the detection and at the start of the yellow marked region, the chemical-potential offset of the whole 1DEC is raised into the zero electron occupation regime (cf. Fig. \ref{fig:fig1}d) such that an electron confined at the left hand side of the 1DEC could tunnel through the transparent barrier there to the SET detected by a current step from 0.06$\,$nA to 0.11$\,$nA (marked in Fig. \ref{fig:fig2}a). The current step height is approximately equal to the one for filling the 1DEC by one electron. The slight difference occurs as the operation point of the SETs is slightly different at loading and unloading due to capacitive cross-coupling with the clavier gates. Most significantly, there are no such SET current step observed during the first D$_L$ segment, which proves the absence of the electron from the emerged left QD and therefore its shuttling. Note that all voltages applied to gates during the first and second D$_L$ are equal, since the shuttling distance is $1\lambda$. After S$_1$, a new QD emerged on the left hand side of the 1DEC. This QD remains unoccupied, since no electron is detected during the first D$_L$, proving that the electron does not tunnel between minima of the propagating wave potential and that the tunnel barrier to the SET is opaque. Using  a multiple of $\lambda$ as a shuttling distance is crucial for the simplicity of our proof, since the operation point of the SET is altered heavily by cross-capacities coupling of the the clavier gates to the sensor QD of the SET giving rise to the current oscillations (following the SETs Coulomb peak of its I-V characteristic) during S$_1$ and S$_{-1}$. Since we shuttled by a distance of $1\lambda$, the sensitivity of the SET is equal during both D$_L$ segments.

So far we discussed a single-shot trace of the shuttling sequence. Its reproducibility becomes obvious from plotting 100 single shot traces recorded by looping the pulse sequence (Fig. \ref{fig:fig2}b). Comparing these traces, detected electron tunnel events become obvious: During the preparation segment, the stochastic nature of the electron tunneling event into the 1DEC is reflected by the duration of the SET current plateau before the current step arises due to charging of the 1DEC. The last detection segment, similarly reveals a stochastic detection of a current step in the opposite direction due to discharging of 1DEC. In the middle, during the first detection segment, a tunnelling event is clearly absent. Thus, the shuttling worked for all single-shot traces. Slight difference between traces other than the stochastic tunnel events are related to slow charge fluctuation of the QuBus device altering the operation point and sensitivity of the SET. In between single shuttling sequences, we slightly correct the operation point of the SET by the voltage applied to LB1 based on its absolute current.

When averaging over the 100 shuttling sequences from Fig. \ref{fig:fig2}b, we identify the stochastic tunneling event as an exponential decrease of the sensor response and increase for P$_1$ and the second D$_L$, respectively (Fig. \ref{fig:fig2}c). The slight increase of current during the first detection segment is assigned to a small transient due to use of bias-tees (see method section).

\begin{figure*}
    \centering
    \includegraphics[width=\linewidth]{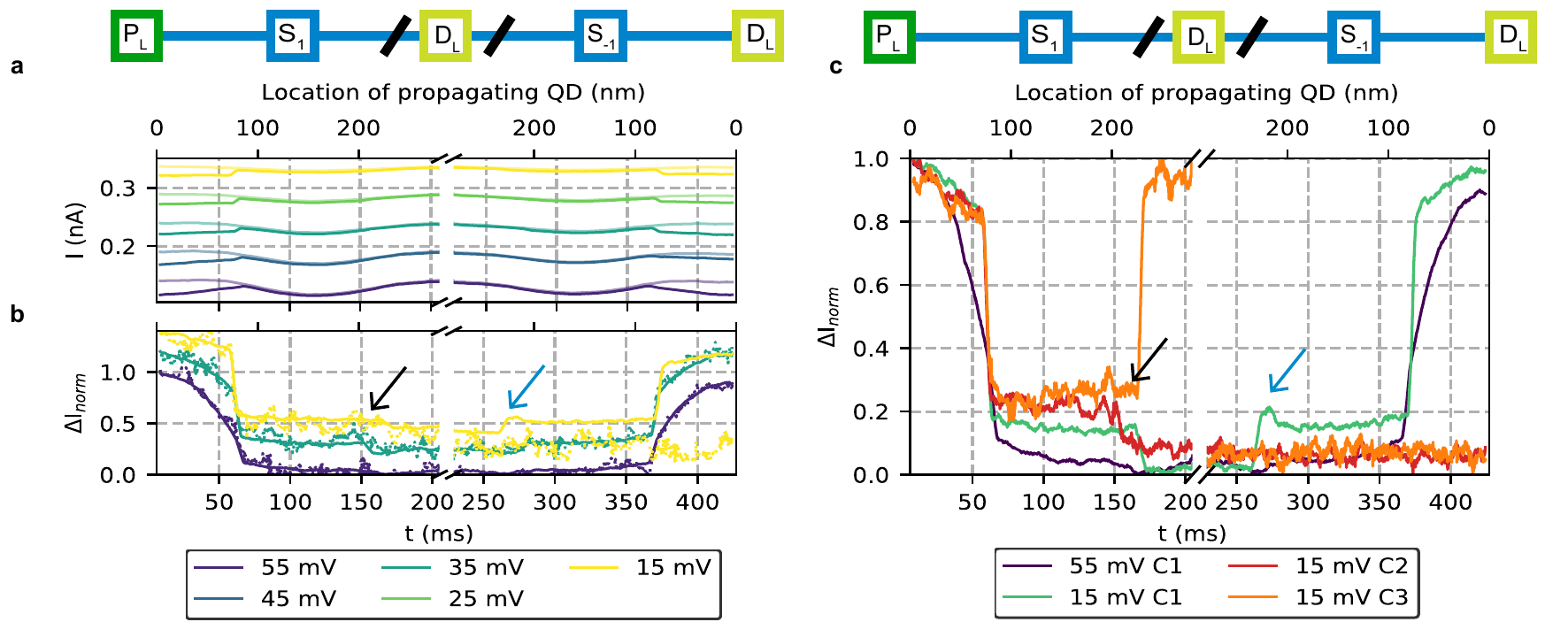}
    \caption{Electron transfer characteristic. The pulse sequences is indicated above the panels. \textbf{a} Comparison of the average shuttling signal with (dark) and without (light, reference signal) an electron loaded for different amplitude $A_S$ values measured by the left SET compensated for cross-capacitive coupling. Shuttling signals  with $B_S=730\,$mV, $\Delta B_S=100\,$mV and $f=5\,$Hz,  are used. For better orientation, the upper x-axis indicates the minimum location of the propagating QD assuming constant shuttling velocity. The curves are offset by 0.05\,nA for clarity. \textbf{b} Normalized difference of average SET currents (solid) and single traces (dotted) with one and without an electron  loaded (from panel a) for different amplitude $A_S$. The curves are offset by 0.1 for clarity. \textbf{c} For $A_S=15\,$mV, 100 traces are bundled into three different categories: the intended transfer (C1) and two different failure modes, with $\Delta I_{norm}=0$ during S$_{-1}$, if the electron is not detected by any D$_L$ (C2) or the electron is detected during the first D$_L$ segment (C3). Each bundle is averaged and normalized separately. The black and blue arrows in panels b and c highlight abrupt changes in $\Delta I_{norm}$.}
    \label{fig:fig3} 
\end{figure*}

Next, we prove that shuttling across the QuBus device is feasible as well. Only three pulse segments are needed: P$_L$,S$_{1.5}$,D$_R$ (Fig. \ref{fig:fig2}d), with P$_L$ being identical to the previous pulse sequence. S$_{1.5}$ shifts the electron by a distance of 1.5$\lambda$, which constitutes the total length of the 1DEC. During D$_R$, the chemical potential of the 1DEC is raised and the tunnel barrier to the SET is set transparent by a voltage pulse applied to the B$_5$ gate. During P$_L$, a current step of the left SET indicates single electron occupation of the 1DEC on the left hand side. During D$_R$, the current step of the right sensor indicates a single electron tunneling out of the 1DEC on its right hand side (As the right SET's operation point is set to a negative flank of a Coulomb peak, the charge unload corresponds to a decrease in current here.). Repeating this pulse sequence 100 times again reveals the stochastic tunneling events in the left SET current during P$_L$ (Fig. \ref{fig:fig2}e) and in the right SET current during D$_R$ (Fig. \ref{fig:fig2}f). Reversely, no current steps are detected by the left SET during D$_R$ and by the right SET during P$_L$. Thus, single shuttling events of single electrons shuttling across the 1DEC are demonstrated. Averaging over the 100 measurement traces (Fig. \ref{fig:fig2}g) reveals the expected exponential decays of the current measured by the left SET for the P$_L$ and for the right SET for the D$_R$ segment. 

Using the detection segments within the proof-of-principle, we demonstrate that a single electron could be shuttled to each end of the 1DEC. Now we discuss whether the proximal charge detectors can provide information of the smoothness of electron shuttling across the 1DEC. It is a characteristic of our conveyor-mode approach that the electron is confined by a smoothly propagating QD. This distinguishes our approach from an electron tunneling across a QD array (bucket-brigade mode). The capacitive cross-coupling of the clavier gates to the sensor QDs of the SET complicates direct observation of the electron movement by an SET. The signals $V_i$ required for shuttling drastically modifies the SET's operation point and diminish its sensitivity. Thus, we compensate for the alteration of the operation point by adjusting the voltages applied to LB1 and RB1 gates during shuttling assuming a linear cross-capacity matrix (virtual gate approach \cite{Nowack2011}). This first measure keeps the sensitivity of the SET within bounds, but does not guarantee constant charge sensitivity during shuttling. Therefore, we interleave the looping of the shuttling pulse sequences (P$_L$,S$_1$,D$_L$,S$_{-1}$,D$_L$) by a nearly identical sequence providing a reference SET trace (Fig. \ref{fig:fig3}a). For this reference, only the P$_L$ segment is modified, such that no electron is loaded. All current variations in the reference trace are thus due to uncompensated capacitive cross-coupling. In Fig. \ref{fig:fig3}a, we focus on the current signal during the two shuttling segments (forth and back cf. Fig. \ref{fig:fig2}a) for various amplitudes $A_S$. Comparing averages of 100 single-shot shuttling traces to their corresponding (zero-electron) reference trace, we conclude that the detection signal of the electron shuttling approaches the one of the reference and merge at $t\approx 70$\,ms, which indicates the electron moving away from the SET. When the electron returns during $S_{-1}$, both traces diverge again. Thus, we detect the decline of the Coulomb interaction between the single electron and the left SET during the shuttling process.

To extract more details from these averaged signal traces, we subtract each reference trace $I_0$ from the electron detection trace $I_1$ and normalize $\Delta I_{norm} = [\Delta I-\min(\Delta I)]/[\max(\Delta I)-\min(\Delta I)] $, where $\Delta I=I_0-I_1$ (Fig. \ref{fig:fig3}b). Strikingly, the averaged curves matches well a randomly picked single trace (dots in Fig. \ref{fig:fig3}b) for  
for $A_S=(35,55)\,$mV. We conclude, that the smoothness of the averaged $\Delta I_{norm}$ curve is thus not only a result of averaging stochastic tunneling events during shuttling, but each single shuttling process itself is a smooth motion as expected for the movement of the QD in the propagating potential. For the smallest amplitude $A_S=15$\,mV, a ripple appears at times marked by arrows and the initial decline is more abrupt. The averaged $\Delta I_{norm}$ at $A_S=15$\,mV does not fully recover after S$_{-1}$. This asymmetry is caused by the electron not returning in every single-shot trace and therefore reducing the average. Therefore, the amplitude $A_S=15$\,mV is not sufficient to confine the electron during the shuttled motion. A threshold for $A_S$ is expected as the confinement of moving QD has to be larger than the potential disorder due to charged defects in the device.

As we record current traces of single shuttling events, we are able to analyze individual failure modes of the electron shuttling at this small amplitude $A_S=15$\,mV. We find two typical failures within the 100 shuttling traces. We bundle these traces into three categories (C1, C2, C3) and separately average them (Fig. \ref{fig:fig3}c). C1 labels shuttling forth and back without failure as confirmed by the D$_L$ segments. For 14\% of traces labeled C2, the electrons shuttles forth, but does not return during $S_{-1}$ and presumably get trapped in the 1DEC. (In some rare cases such electrons become unloaded during the reference shuttling sequence, prior to resetting 1DEC charge state of the flush pulse sequence described in Methods). For 5\% of traces labeled C3, the shuttling failed already during $S_{1}$, at $t \approx 160$\,ms (black arrow), since the detection signal suggest the electron tunnels back to left end of the 1DEC. Also during $S_{-1}$ at $t \approx 270$\,ms (blue arrow), the traces labelled C1 show a ripple indicating a tunnel event. Note that the two arrows mark in fact the same position in the 1DEC, assuming the shuttling velocity is constant. Presumably, potential disorder in the 1DEC poses an unintentional barrier there. For $A_s=55$\,mV, all 100 single-shot shuttling traces show no failure (C1) and abrupt steps in current are absent confirming smooth shuttling.  The decline of $\Delta I_{norm}$ lasts till $t \approx 170$\,ms corresponding to a shuttling distance of 230\,nm and remains continuous underlining the smoothness of the electron transfer. We conclude that the intentional propagating potential in the 1DEC provides enough confinement at $A_s=55$\,mV exceeding the potential disorder. Similarly, at $t\approx 67$\,ms and correspondingly at $t\approx 366$\,ms the detection curves show increasing smoothness as  $A_s$ is increased (Fig. \ref{fig:fig3}b,c). This analysis demonstrates that the single-electron shuttling also provides a new method for mapping of electrostatic disorder.

\begin{figure}
    \centering
    \includegraphics[width=8cm]{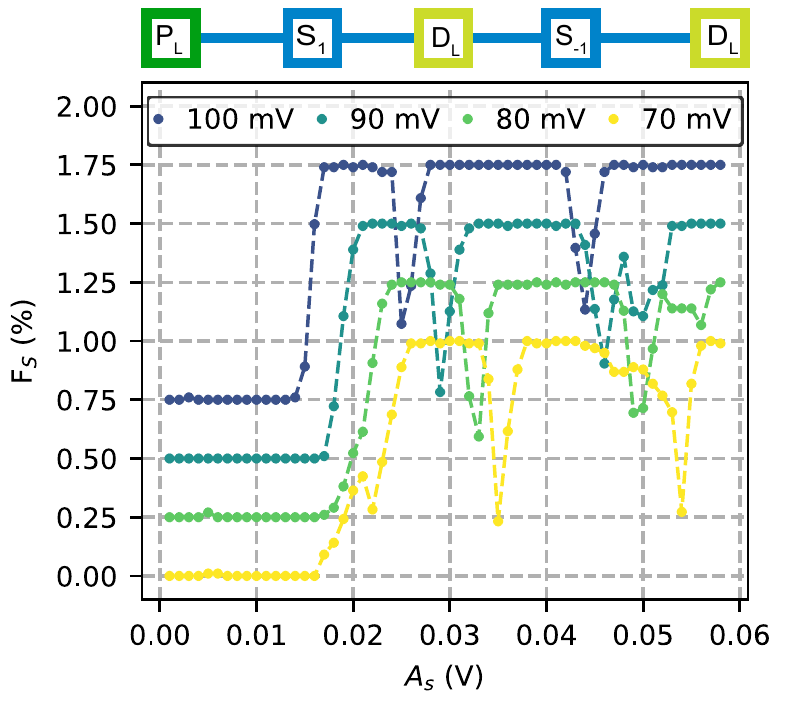}
    \caption{Shuttle fidelity as a function of shuttling amplitude for different voltage offsets $\Delta B_S$ between the two different gate layers (P and B gates) with fixed $f=1$\,kHz and $B_S=730$\,mV. The pulse sequence is depicted above. The electron is loaded on the left, shuttled to the right by a distance of $\lambda$, a detection segment is applied, the electron is shuttled back to the left and a second detection segment is applied. We capture the success probability $P$ as a function of the shuttling amplitude $A_s$ and different offsets between the gate layers $\Delta B_s$. The shuttling sequence is counted as successful if the first D$_L$ detects no electron and the second one does detect an electron.  The curves are offset by 0.25 for clarity and solid lines are guides-to-the eye.}
    \label{fig:fig4} 
\end{figure}

We complete the failure mode analysis by measuring the shuttling fidelity $F_S$ as a function of the amplitude $A_S$ and offset $\Delta B_S$ of the $V_i$ signals. This analysis confirms that $A_S$ higher than a certain threshold allows for single electron shuttling with high $F_S$ for the shuttling scheme (P$_L$,S$_1$,D$_L$,S$_{-1}$,D$_L$) looped 100 times (Fig. \ref{fig:fig4}). We define a shuttling attempt as successful, if zero electron detection during the first D$_L$ coincides with one electron detection during the second D$_L$ segment within one shuttling trace. Here, we apply a charge detection scheme with a detection infidelity of $2\cdot10^{-5}$ (see Methods). $F_S$ extracted from 100 measurement repetitions exhibits two distinct features: (I) When increasing the shuttling sine wave amplitude $A_s$, the first shuttling events are observed starting at approximately $A_s=14$\,mV. Once $A_s$ is sufficiently large, the shuttling fidelity $F_s$ mostly remains beyond $99\%$. This is understandable, since the expected potential disorder within the 1DEC due to charged defect has to be small compared to shutting potential. Thus, we confirm that larger $A_s$ and thus stronger electron confinement by the propagating wave supports smooth single electron shuttling. It also matches our observation of several failure modes at $A_s=15$\,mV (cf. Fig. \ref{fig:fig3}c). (II) The $F_s$ drops significantly for some amplitudes. These dips in $F_S$ alter systematically with an increase in the offset $\Delta B_S$. The characteristic dependence of the $F_S$ dip as a function of ($A_S, \Delta B_S$) suggest a resonance with a charge defect trapping the electron. By choosing appropriate ($A_S, \Delta B_S$), the operation of the QuBus is ensured. Note that thermal cycling the QuBus alters and can even remove these dips. We observe that shuttling the electron back and forth to 5000 times before charge detection readout does not decrease $F_S$. A separate measurement at $A_s = 50\,$mV, $\Delta B_s= 100\,$mV and $f =1\,$ kHz indicated a fidelity $F_S = 99.42\pm0.02 \%$ using 150000 single electron shuttling sequences. Note this $F_S$ includes infidelity of the segment $L_L$, which we have not measured separately.          

Our proof-of-principle demonstrates that conveyor-mode shuttling of a single electron in an electrostatically defined quantum channel in Si/SiGe is feasible. The four input signals $V_i$ controlling the travelling potential are parameterized by $A_S, B_S, \Delta B_S, \varphi(t)$ and can be simply tuned. Besides setting roughly the chemical potential of the quantum channel by $B_S$, $\Delta B_S$, most importantly, the amplitude $A_S$ has to be sufficiently large to confine the electron despite unintentional potential fluctuations distributed along the quantum channel. For our device a moderate amplitude is sufficient to shuttle across the 420\,nm long channel with fidelity exceeding 99\%. The phase $\varphi(t)$ sets the electron shuttling velocity and de/acceleration, including the demonstrated reversal of the shuttling direction. The time-resolved comparison of a shuttling pulse sequence with none and a single electron indicates a smooth electron motion expected for the conveyor-mode. By continuous monitoring of single shuttling events, failure modes can be categorized, which provides new means to localise critical potential disorder. The smoothness of the electron motion will be an important parameter for modelling spin-coherent shuttling of a spin qubit. Specifically, we expect the shuttling distance, velocity and acceleration has to balance orbital adiabaticity on the one hand and the spin dephasing and relaxation time of a static quantum dot on the other hand \cite{Huang2013, zhao2016,Hollmann2020, Struck2020}. An extension of the QuBus length is possible without additional input signals. A QuBus length of 1 to 10 micron gains sufficient space between dense qubit registers to interleave signal vias and control electronics tiles. For the latter, the simple control signals required for our QuBus do not need local memory on the control electronic tiles and thus ease the integration of the QuBus into a scalable quantum computing architecture.

\section{Methods}

\label{sec:methods}
\textbf{Setup and measurement procedures}:
All experiments are executed in a dilution refrigerator with a base temperature of 40$\,$mK. All dc lines to the device are filtered by pi-filters ($f_c=5\,$MHz) at room temperature and by 2nd order RC filters with $f_c=10$\,kHz at base temperature. The clavier gates B3, P4 and B5 are connected to resistive bias-tees with a cutoff frequency of 5$\,$Hz. Signals are applied the ac and dc input terminal of the bias-tee, in order to effectively neutralize the bias-tees, since low-bandwidth pulses are required for the presented experiments. A serial resistor is added to the low-frequency terminal, the value of which is tuned by flattening sensor signal response. Electrical connections among clavier gates (Fig. \ref{fig:fig1}a) are wired up outside the cryostat for flexibility reasons. The SETs are dc-biased by 500\,$\mu$V and readout by a transimpedance amplifier and an analogue-digital-converter.

\textbf{Flush pulse sequence}: To ensure an electron-free 1DEC, we apply a flush pulse-sequence to the gates before looping shuttling pulse sequences. The voltages applied to all clavier gates (B$_i$ and P$_i$) are altered by 100$\,$mV in the following sequence: first we decrease the voltage applied to gate B$_3$, then the one applied to P$_2$ and P$_3$, followed by a decrease on B$_2$ and B$_4$, and next on P$_1$ and P$_4$. The flush pulse segment is finished  by increasing the voltages applied to gates B$_1$ and B$_5$ by 100$\,$mV and finally resetting to the original voltages applied to the gates in reverse order. 
The data in Fig. \ref{fig:fig3} and \ref{fig:fig4} was taken by concatenating the pulse sequences for different amplitudes. To ensure that the 1DEC remains electron free throughout the measurement, additional pulses, shuttling potentially left behind electrons out of the 1DEC (S$_{-1}$+D$_L$), are added in between the pulse sequences for different amplitudes.

\begin{figure}
    \centering
    \includegraphics{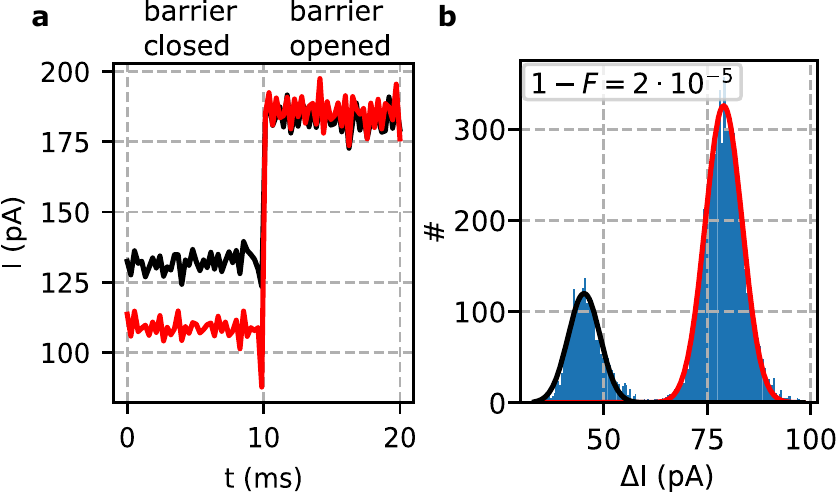}
    \caption{Fidelity $F$ of single-shot charge-readout at the ends of the shuttle device by the SETs. \textbf{a} SET current trace $I$ during the two stages of the detection segment with an electron present (red) and without an electron (black) at the end of the one-dimensional QuBus channel. During the first 10\,ms, the tunnel barrier between the 1DEC and the SET is fully opaque and then fully transparent during the second 10$\,$ms allowing an electron potentially present to immediately leave the 1DEC. \textbf{b} Histogram of the current steps $\Delta I$ between the two detection stages in panel a for 11300 traces with arbitrary charge state 0 or 1 at the end of the channel before the start of the detection segment. The events covered by the two fitted Gaussians are assigned to the electron occupation 0 and 1 (red and black Gaussian fit), respectively.
    }
    \label{fig:fig5}
\end{figure}

\textbf{Single-shot charge detection}: There are two methods applied for the detection segment D$_x$ using the proximal SETs on either the left or the right end of the QuBus device: (I) The tunnel rate between reservoir and first QD of the 1DEC is raised corresponding to the duration of the detection segment, which is long compared to the SET bandwidth. Thus, a single electron tunneling event from the first QD of the one dimensional channel can be resolved by a step in the current across the SET, as the absence of the electron alters the operation point of the SET.This method is applied for the measurements presented in Fig. \ref{fig:fig2}. (II)  The second charge detection method employs a two-stage SET current-readout each 10\,ms long. During the first stage, the tunnel barrier between the 1DEC and the SET is opaque and then made fully transparent by pulsing the voltage applied to the barrier gate during the second stage, i.e. the tunnel rate is faster than the measurement bandwidth of 1\,kHz. This results in a difference between the SET current levels measured during the two stages, which depends on the electron occupation of the QD at the end of the 1DEC before the detection segment (Fig. \ref{fig:fig5}a). In the absence of the electron, this current step is determined by the cross-coupling of the barrier gate to the SET's QD alone. Plotting the observed current steps $\Delta I$ for 10000 detection segments, we observe a distribution fitted by two Gaussians (Fig. \ref{fig:fig5}b) assigned to the two distinct readout results. Calculated from the overlap of the two Gaussian's, the detection infidelity is $1-F=2\cdot10^{-5}$. This second method is applied for the measurements presented in Fig. \ref{fig:fig3} and \ref{fig:fig4}.   

The data discussed was obtained after a thermal cycle to 100$\,$K. Previously, a similar behaviour of the sample was observed. Only the values of the dc voltages were retuned.

\textbf{Device fabrication}: Fig. \ref{fig:fig1}a. depicts the SEM inspection of a device that is fabricated identically to the measured one. It is fabricated on an undoped strained Si/SiGe heterostructure utilizing the metal lift-off technique. A 10$\,$nm thick strained Si layer acts as the quantum well covered by a 30$\,$nm undoped Si$_{0.7}$Ge$_{0.3}$ spacer and a 2$\,$nm Si cap layer. Ohmic contacts to the quantum well layer are selectively implanted by phosphorus ions and activated by rapid thermal processing at 700$\,$°C for 15$\,$s. A combination of electron beam lithography, metal evaporation, metal lift-off process, and atomic layer deposition of dielectrics enables the overlay of a three layers of electrically isolated metallic gates, where Ti and Pt are gate metals insulted by 10$\,$nm Al$_2$O$_3$ from the substrate and among gate stacks. Each layer of metal gates consists of 5 $\,$nm Ti as the adhesion metal and a varied thickness of Pt: 15 $\,$nm, 22 $\,$nm and 29 $\,$nm for layer 1 to layer 3 correspondingly, which robustly ensures the continuity of metal gates in a three layers stack despite the existence of an inhomogeneous local topography. The two gates colored in purple in Fig. \ref{fig:fig1}a are geometrically separated by 200 $\,$nm providing a 1DEC for single electron shuttling. In this experiment, a device with 60 $\,$nm clavier gate width is measured. The clavier gates are separated by 10$\,$nm, thus the gate pitch is $g=70$\,nm. The SETs employed as charge sensors or as electron reservoirs are formed by gates distributed among two gate layers. The charge carriers in the transport channel are accumulated by a top-gate (yellow gates in Fig. \ref{fig:fig1}a.) with barrier gates (LB1, LB2, RB1, RB2) lying underneath.  The sensor dots can be formed at both ends of the 1DEC symmetrically.

\section{Acknowledgements}
The authors thank Jürgen Moers, Lieven Vandersypen and Anne-Marije Zwerver for useful discussion. This work has been funded by the German Research Foundation (DFG) under Germany's Excellence Strategy - Cluster of Excellence Matter and Light for Quantum Computing" (ML4Q) EXC 2004/1 - 390534769 and by the Federal Ministry of Education and Research under Contract No. FKZ: 13N14778. Project Si-QuBus received funding from the QuantERA ERA-NET Cofund in Quantum Technologies implemented within the European Union's Horizon 2020 Programme. The device fabrication has been done at HNF - Helmholtz Nano Facility, Research Center Juelich GmbH.\cite{albrecht_hnf_2017}

\section{Author contributions}
I.S. and T.S. did the shuttling measurements and the data analysis supported by R.X. and L.R.S..  I.S. and R.X. fabricated and pre-characterized the device. S.T. did electron-beam lithography. N. F. did the electrostatic simulations. L.R.S. conceived and supervised the study supported by H.B. and all authors discussed the results. T.S., I.S., R.X. and L.R.S. wrote the manuscript, which all other authors reviewed.

\section{Competing interests}
The authors declare no competing interests.

\section{Data availability}
The datasets generated during and/or analysed during the current study are available from the corresponding author on reasonable request.

\bibliography{citations}

\end{document}